\documentclass[11pt,letterpaper]{article}
\usepackage[T1]{fontenc}
\usepackage[utf8]{inputenc}
\usepackage{lmodern,amsmath,amssymb,amsthm,mathtools}
\usepackage[margin=1in]{geometry}
\usepackage{microtype,booktabs,array,enumitem,tikz}
\usepackage[hidelinks]{hyperref}
\hypersetup{pdftitle={Beyond the Quadratic Barrier for Multidimensional Grid Range Problems},pdfauthor={Filippo di Gravina and Francesco Damiano Di Gregorio}}
\usepackage{xurl}
\usetikzlibrary{arrows.meta,positioning}
\setlist{nosep}
\newtheorem{theorem}{Theorem}
\newtheorem{lemma}[theorem]{Lemma}
\newtheorem{corollary}[theorem]{Corollary}
\newtheorem{proposition}[theorem]{Proposition}
\newtheorem{assumption}[theorem]{Assumption}
\newcommand{\Ot}{\widetilde O}
\newcommand{\ind}{\mathbf 1}
\newcommand{\chmax}{\operatorname{chmax}}
\newcommand{\chmin}{\operatorname{chmin}}
\newcommand{\add}{\operatorname{add}}
\newcommand{\set}{\operatorname{set}}

\newcommand{\polylog}{\operatorname{polylog}}
\title{Beyond the Quadratic Barrier for Multidimensional Grid Range Problems}
\author{Filippo di Gravina \qquad Francesco Damiano Di Gregorio\\
{\normalsize\textit{Email:} \href{mailto:f.digravina2@studenti.uniba.it}{\texttt{f.digravina2@studenti.uniba.it}},\quad
\href{mailto:f.digregorio16@studenti.uniba.it}{\texttt{f.digregorio16@studenti.uniba.it}}}}
\date{November 30, 2026}
\begin{document}
\maketitle
\begin{abstract}
We give truly subquadratic algorithms for two-dimensional Grid Range problems with sum queries and non-distributive updates. For $\chmax$ updates, we obtain deterministic $N^{22/13+o(1)}$ offline time under a square matrix multiplication exponent of $9/4$, and $\Ot(N^{7/4})$ combinatorial time even online, addressing an open problem of Lau and Ritossa (ITCS 2021). For mixed additions and assignments we obtain $N^{(25-\sqrt{19})/12+o(1)}$ offline time under an explicit rectangular arithmetic assumption, and $\Ot(N^{7/4})$ combinatorially. The common technique is to compress local histories in a spatial partition rather than materialize full-cover updates. The resulting trellised problems are equivalent to weighted masked products of two and three sparse matrices, respectively. We also obtain a matching conditional online local lower bound, a fixed-dimensional static maximum-query classification, and $\Ot(N^{(d+1)/2})$ time for static sums with additions and assignments.
\end{abstract}

\section{Introduction}
Grid Range problems ask for aggregate queries on an implicit grid subject to box updates. Linear operations admit efficient range-searching techniques, but assignment and saturation updates can overwrite contributions on an enormous number of grid points. Lau and Ritossa~\cite{lr} asked whether two-dimensional sum queries with $\chmax$ updates admit truly subquadratic time (Open Problem 8.3, or 52 in the full version). Their mixed addition/assignment algorithm also has a substantially larger exponent than the bounds below.

Our approach is to decompose the grid into trellised cells, where every partial update is a strip, and compress the effect of full-cover updates. For $\chmax$, dominated fragments are discarded and the surviving one-dimensional histograms have short signed histories. For mixed additions and assignments, a last-writer identity gives a sparse chronological join. Static queries admit a further simplification: both sums and maxima reduce to one-dimensional marginal profiles.

\begin{table}[ht]
\centering\small
\begin{tabular}{@{}lllcc@{}}\toprule
Query & Updates & Model & Algebraic exponent & Combinatorial exponent\\\midrule
$\sum$ & $\chmax,\ \chmin,\ \set$ (separate) & offline 2D & $22/13$ & $7/4$\\
$\sum$ & $\add,\set$ (mixed) & offline 2D & $(25-\sqrt{19})/12$ & $7/4$\\
$\sum$ & $\chmax,\ \chmin,\ \set$ (separate) & online $d$D & -- & $d/2+3/4$\\
$\sum$ & $\add,\set$ (mixed) & static $d$D & -- & $(d+1)/2$\\
$\max$ & $\add,\set,\chmax$ & static $d$D & -- & $(d+1)/2$\\
$\max$ & $\set,\chmin,\chmax$ & static $d$D & -- & $(d+1)/2$\\\bottomrule
\end{tabular}
\caption{Deterministic total-time upper bounds, suppressing subpolynomial or polylogarithmic factors. Static means that all queries follow all updates. Algebraic bounds use Assumption~\ref{ass:arithmetic}.}
\label{tab:bounds}
\end{table}

The improvement is structural: even at the same arithmetic envelope, the previous assignment bound $5/4+\omega/(\omega+1)$ decreases by $1/4$ to $1+\omega/(\omega+1)$. Likewise, the previous mixed-update framework, reoptimized at our rectangular envelope, yields $(28-\sqrt{19})/12\approx1.9701$, compared with our $1.7201$. The general dynamic maximum problem with mixed addition and assignment (Open Problem 50 of~\cite{lr}, full version) remains open.

\paragraph{Model.} The grid $[s]^d$ is initially zero. There are $N$ box updates and queries in total, $d$ is fixed, and $s$, coordinates and signed integer weights are polynomially bounded in $N$. The maps are $v\mapsto v+w$, $v\mapsto w$, $v\mapsto\max(v,w)$ and $v\mapsto\min(v,w)$. We work on a word RAM with constant-word arithmetic; represent points by unit cells and boxes by half-open intervals. All lengths and volumes count actual grid points. Offline algorithms know the entire sequence; online algorithms answer before the next operation. We write $\Ot$ for polylogarithmic factors.

\begin{assumption}\label{ass:arithmetic}
For fixed $a,b\ge0$, exact arithmetic multiplication of dimensions $N^a\times N^b\times N^a$ has exponent at most
\[
\mu(a,b)=\begin{cases}2a&b\le a/2,\\ a+b+a^2/(4b)&b\ge a/2.\end{cases}
\]
We stipulate the rectangular curve of Uzna\'nski~\cite{uz}, including $\omega\le9/4$, and straight-line implementations over suitable finite fields or constant-degree extensions, with exact reconstruction of polynomially bounded signed answers.
\end{assumption}
Combinatorial bounds do not use this assumption. For arithmetic algorithms, exponent bounds include arbitrary fixed slack (written $o(1)$); field size or CRT moduli suffice to recover exact answers.

\section{Two common tools}
\subsection{Trellised decomposition and full-cover histories}
\begin{lemma}[Spatial partition]\label{lem:space}
For $N$ rectangles and $B=\lceil\sqrt N\rceil$, a spatial tree with $O(N)$ leaves of depth $O(\log N)$ can be built in $\Ot(N^{3/2})$ time such that every operation decomposes into $\Ot(B)$ disjoint full nodes and leaf strips; each leaf has $O(B)$ partial intersections. In fixed dimension $d$, the corresponding bounds are $O(N^{d/2})$ leaves, $O(B)$ partial intersections per leaf and $\Ot(N^{(d+1)/2})$ total pieces.
\end{lemma}
\begin{proof}
In 2D, cut the sorted multiset of vertical-side coordinates at every $B$th occurrence, merging equal cuts; do the same for horizontal sides. Within each x-slab also cut at the two y-endpoints of every rectangle having an interior vertical side. There are $O(N/B)$ slabs, each with $O(N/B+B)$ cells. No rectangle corner is strictly inside a cell, so each nonempty intersection is full or a strip. Every partial intersection has an interior side, and at most $O(B)$ side occurrences of either orientation lie inside a cell. In each slab, build a balanced binary tree on its y-cells; connect slab roots by a balanced tree. In fully x-contained slabs use canonical y-nodes and at most two partial leaves; enumerate cells in the at most two boundary slabs. This costs $\Ot(N/B+B)$ per operation. In $d$ dimensions use the Overmars--Yap partition~\cite{oy,lr} (full version, Lemma 33 and Theorem 21). Refine each penultimate slab's last-axis child list into a balanced tree. Scan the $O(N^{(d-1)/2})$ penultimate slabs: when fully contained in the first $d-1$ axes, use canonical last-axis nodes, otherwise enumerate the intersected children, all partial. The imported bound of $O(N^{(d-1)/2})$ partial leaves per box gives the claim.
\end{proof}

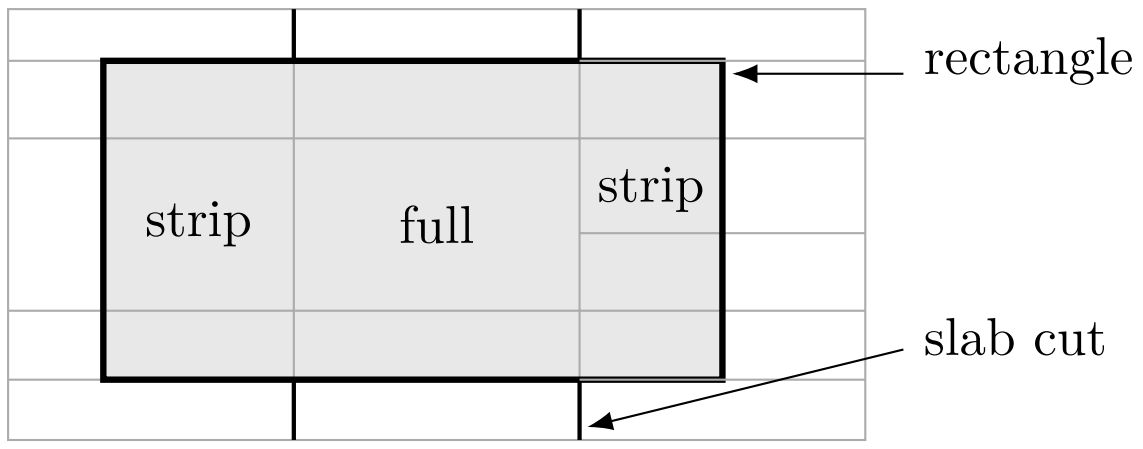
\begin{figure}[ht]
\centering
\begin{tikzpicture}[x=0.63cm,y=0.57cm,font=\small]
\draw[gray!65] (0,0) rectangle (9,5);
\foreach \x in {3,6} \draw[thick] (\x,0)--(\x,5);
\foreach \y in {1.5,3.5} \draw[gray!65] (0,\y)--(9,\y);
\draw[gray!65] (0,0.7)--(3,0.7) (0,4.4)--(3,4.4) (6,2.4)--(9,2.4);
\fill[gray!18] (1,0.7) rectangle (7.5,4.4);
\draw[gray!65] (3,0.7)--(3,4.4) (6,0.7)--(6,4.4) (1,1.5)--(7.5,1.5) (1,3.5)--(7.5,3.5) (6,2.4)--(7.5,2.4);
\draw[very thick] (1,0.7) rectangle (7.5,4.4);
\draw[gray!65] (6,0.7)--(9,0.7) (6,4.4)--(9,4.4);
\node at (4.5,2.5) {full}; \node at (2,2.5) {strip};
\node at (6.75,2.9) {strip};
\node[anchor=west] at (9.4,4.4) {rectangle};
\draw[-{Latex}] (9.4,4.25)--(7.6,4.25);
\node[anchor=west] at (9.4,1.2) {slab cut};
\draw[-{Latex}] (9.4,1.05)--(6.08,0.15);
\end{tikzpicture}
\caption{A rectangle is represented by full canonical nodes and partial strips; the extra horizontal cuts remove interior corners.}
\end{figure}

Let $S$ denote the total geometric size, so $S=\Ot(N^{(d+1)/2})$, and $k_C=1+$ the number of partial intersections of leaf $C$. Thus $k_C=O(B)$ and $\sum_C k_C=O(S)$. In 2D, a local cost $\Ot(k_C^c)$ lifts to $\Ot(N^{1+c/2})$, provided global aggregates and full-cover histories are handled in $O(S\polylog N)$ time. The proviso is essential: explicitly sending every full update to every covered leaf can cost quadratic time.

\begin{lemma}[Gap compression]\label{lem:gap}
Suppose full-cover update maps form a constant-size representation closed under composition, and each map is realizable by $O(1)$ local updates. All full-cover gaps between consecutive partial operations in every leaf can be compressed into $O(k_C)$ synthetic updates per leaf in $\Ot(S)$ total time.
\end{lemma}
\begin{proof}
Build a sparse lazy spatial tree for each canonical interval of a binary tree on time. Insert each full-node update into the $O(\log N)$ time ancestors of its timestamp, in chronological order. Compose new maps after old tags, pushing ancestors before descending; allocate only touched paths. Total construction is $\Ot(S)$. For a leaf and a gap, decompose the time interval into canonical intervals in chronological order. Within each corresponding sparse spatial tree, compose tags on the leaf path, descendants before pending ancestors. This gives exactly the full-cover transform during that interval; concatenate the transforms in chronological order. Include the initial and, when needed, final gaps. Each gap takes polylogarithmic time and there are $O(S)$ gaps. Full-cover maps commute neither with partial updates nor with each other in general; this construction preserves their order.
\end{proof}

\subsection{Sparse masked arithmetic joins}
We use a single comparison encoding throughout. For a strict comparison $a<b$, take a balanced binary tree on the sorted ranks. The smaller operand carries labels of ancestors at which it lies in the left subtree, and the larger operand labels at which it lies in the right subtree. They share exactly one label, their lowest common ancestor, iff the comparison holds. For $a\le b$ add an equality label. Each operand has logarithmically many labels. Independent comparison types use disjoint label spaces.

\begin{lemma}[Masked products]\label{lem:products}
Let $M$ count the nonzeros of sparse matrices and the number of requested entries. Masked $AB$ is computable in $M^{2\omega/(\omega+1)+o(1)}$ arithmetic time; masked $ABC$ is computable in $M^{c+o(1)}$ time, $c=(13-\sqrt{19})/6$, under Assumption~\ref{ass:arithmetic}. Both admit deterministic combinatorial $\Ot(M^{3/2})$ algorithms, including signed weights.
\end{lemma}
\begin{proof}
For $AB$, view queries as sparse dot products of rows and columns. Scan the smaller operand whenever it has at most $D$ entries. Among the remaining $O(M/D)$ rows and columns, enumerate all pairs of incidences for labels of total degree at most $D$. Both steps cost $\Ot(MD)$. There remain $O(M/D)$ heavy labels, so one square multiplication costs $(M/D)^{\omega+o(1)}$. Balance at $D=M^{(\omega-1)/(\omega+1)}$. Combinatorially, take $D=\sqrt M$ and scan the heavy labels per requested pair.

For $ABC$, append a fourth edge layer (the mask), with independent variables $z_{ij}$ on its edges. The sum of products of weights of all cycles traversing the four layers is $F(z)=\sum_{ij}z_{ij}(ABC)_{ij}$. Thus all requested entries are obtained by reverse differentiation at $z=1$~\cite{bs}; topology-dependent enumeration and exact arithmetic gates suffice. For the combinatorial circuit, order vertices by degree, breaking ties by identifier. For each vertex $v$, enumerate paths $v,u,w$ with both other vertices smaller than $v$ and $w$ in its opposite layer. For each $(v,w)$ sum the products separately through each intermediate layer, and multiply the two sums. Each layered cycle is counted at its unique largest vertex. The scan cost is $\sum_{uv\in E}\min(\deg u,\deg v)=O(M^{3/2})$: edges with an endpoint of degree at most $\sqrt M$ cost at most $M\sqrt M$, and the remaining edges are charged to their smaller endpoint, with at most $2\sqrt M$ high neighbors per vertex. Deterministic sorting aggregates the sums. Reverse-mode differentiation has constant-factor arithmetic overhead (for $z=ab$, propagate $b\bar z$ to $a$ and $a\bar z$ to $b$). This adapts the degree-ordered cycle counting approach of~\cite{bh}.

For the algebraic circuit, split vertices into high (degree $>\Delta$), medium ($\sqrt\Delta<\deg\le\Delta$), and small ($\deg\le\sqrt\Delta$). Call medium and small low. Evaluate separately the constant number of layer-class patterns:
\begin{itemize}
\item Two opposite low vertices: contract through both by wedge enumeration and take the dot product of the resulting sparse matrices.
\item At most one low vertex: contract through it (or any vertex if all high) and its opposite, using dense multiplication through high vertices.
\item Two adjacent low vertices: if both are small, enumerate paths around their connecting edge and test the closing edge. Otherwise choose a medium vertex as an outer endpoint, contracting through the other low vertex and its opposite high vertex.
\end{itemize}
Every enumeration costs $\Ot(M\Delta)$, and the largest multiplication has dimensions $M/\Delta,M/\Delta,M/\sqrt\Delta$ in some order. Reverse differentiation again returns the mask. With $\Delta=M^x$, tensor-leg permutation and Assumption~\ref{ass:arithmetic} give
\[
\max\left\{1+x,\;2-\frac32x+\frac{(1-x)^2}{4(1-x/2)}\right\}.
\]
Balancing the increasing and decreasing branches yields $x=(7-\sqrt{19})/6$, hence $c=1+x$. This extends the contraction analysis in~\cite{lr} to signed weights. Polynomial input magnitudes and constant walk length permit exact reconstruction.
\end{proof}

\section{Sum queries with monotone updates}\label{sec:mono}
\begin{theorem}\label{thm:mono}
Offline 2D Grid Range $(\sum,\chmax)$ admits deterministic $N^{1+\omega/(\omega+1)+o(1)}$ time, hence exponent $22/13$ under Assumption~\ref{ass:arithmetic}, and $\Ot(N^{7/4})$ combinatorial time. The same holds for $\chmin$ and $\set$ separately.
\end{theorem}
\paragraph{Canonical fragments.} Use Lemma~\ref{lem:space}. At each leaf make two balanced axis trees, using endpoints of partial operations. Store one positive weight or no weight at each spatial node and each canonical axis node. A point's value is the maximum of zero and the weights on its spatial and two axis paths. Maintain the invariant that every positive stored weight strictly exceeds all positive ancestors in its own tree and in the spatial hierarchy; the two axis trees are incomparable.

To insert $w>0$ at a canonical node, discard it if an ancestor or the node itself has weight at least $w$. Otherwise delete the old weight and every descendant weight at most $w$, then insert $w$. The update covers the deleted fragments and dominates them forever. Subtree minima of positive weights find deletions by logarithmic-depth searches. Each deletion consumes an insertion, so spatial modifications cost $\Ot(N^{3/2})$ altogether, and the strip fragments of each leaf change only $\Ot(k_C)$ times, even when a full spatial update causes deletions.

\paragraph{Intrinsic histograms.} For an axis node $v$, let $L_v$ be its length, $A_v$ the length covered by positive weights in its subtree, and $H_v$ the histogram of intrinsic maxima ignoring ancestors. Then
\[
A_v=\begin{cases}L_v&v\text{ stores a weight},\\\sum_{u\text{ child of }v}A_u&\text{otherwise},\end{cases}\qquad
h_v=L_v-\sum_{u\text{ child of }v} A_u.
\]
A positive node of weight $w$ contributes mass $h_v$ at $w$; $H_v$ consists of all such contributions in its subtree plus mass $L_v-A_v$ at zero. This follows from strict domination, since descendants win wherever covered. A fragment modification changes $A_v,h_v$ only along its ancestor path. Send each changed contribution as a signed mass record to all its canonical ancestors, correcting their zero masses by the opposite net amount. Each modification yields $O(\log^2 N)$ records. Thus all histograms of one leaf have $\Ot(k_C)$ records including initialization.

\paragraph{Full sums.} Let $X,Y$ be the two root histograms. Within each weight range a balanced weight tree stores total masses $X'_0,Y'_0$, first moments $X'_1,Y'_1$, and $M=\sum_{u,w}\max(u,w)X(u)Y(w)$. At a singleton $w$, $M=wX(w)Y(w)$; merging lower and higher children gives
\[
M=M_L+M_R+X'_{0,L}Y'_{1,R}+Y'_{0,L}X'_{1,R}.
\]
Hence root-histogram changes cost logarithmic time. The intrinsic leaf sum is $M$ and positive area is $L_x L_y-X(0)Y(0)$. At a spatial node $v$ of area $S_v$, let $(A',V')$ be its children's total covered area and intrinsic sum (at a leaf, use the strip statistics). With own weight $w>0$, set $(A_v,V_v)=(S_v,V'+w(S_v-A'))$; without a weight, use $(A',V')$. A full-node query with ancestor maximum $a$ returns $V_v+a(S_v-A_v)$. Strict domination guarantees these formulas. Every modification or full-node query costs polylogarithmic overhead beyond fragment deletions.

For a partial query along x, decompose its interval into canonical nodes $v$. With $t$ the largest spatial or proper axis ancestor weight (including zero), each summand is
\begin{equation}\label{eq:phi}
\Phi(H_v,Y,t)=\sum_{u,w}\max(t,u,w)H_v(u)Y(w).
\end{equation}
Record these evaluations with their query times. The number in each leaf is $\Ot(k_C)$; the simulation so far costs $\Ot(N^{3/2})$.

\begin{lemma}[Histogram histories]\label{lem:hist}
Given $P$ signed histogram mass changes and $Q$ requests of the form~\eqref{eq:phi} at specified times, all answers reduce in near-linear time and size to masked $AB$ and ordinary orthogonal prefix sums.
\end{lemma}
\begin{proof}
At time $\tau$, let $H,K$ be the two histogram prefix sums. Partition the pairs into three domains:
\[
\Phi=tH(\le t)K(\le t)
+\!\sum_{u>t,\,u\ge w}uH(u)K(w)
+\!\sum_{w>t,\,w>u}wH(u)K(w).
\]
The first term needs two orthogonal prefix sums on (change time, weight). For each histogram build canonical range subsets of its records; a restriction to time $\le\tau$ and optionally weight $>t$ is a disjoint polylogarithmic collection of subsets. Every record occurs in only polylogarithmically many subsets. Store each subset as a sparse vector on the comparison labels of Section 2.2, summing signed masses multiplied by their weights on the winning side and unweighted masses on the losing side. Use non-strict labels for the second term and strict labels for the third. Expand each query into masked dot products between its two subset lists. Uniqueness of the comparison label and disjointness of the subsets count every eligible record pair exactly once; signed changes cancel by bilinearity. The vectors and mask have size $\Ot(P+Q)$.
\end{proof}
\begin{proof}[Proof of Theorem~\ref{thm:mono}]
Apply Lemmas~\ref{lem:hist} and~\ref{lem:products} to each leaf, with $P+Q=\Ot(k_C)$. For $\gamma=2\omega/(\omega+1)$ or $3/2$ combinatorially, the total is $\Ot(\sum_C k_C^\gamma)=N^{1+\gamma/2+o(1)}$ (polylogarithmic slack in the combinatorial case). Negation handles $\chmin$. For assignments, following~\cite{lr}, maintain two $\chmax$ instances, with weights $Ct+w_t$ and $Ct$ for an assignment of $w_t$ at time $t$, where $C>2\max_t|w_t|$. Both encodings select the last assignment, so the difference of the sums is exact. The same reduction is online with a fixed polynomial word bound.
\end{proof}

\section{Mixed additions and assignments}\label{sec:mixed}
\begin{theorem}\label{thm:mixed}
Offline 2D Grid Range $(\sum,\{\add,\set\})$ admits deterministic $N^{(25-\sqrt{19})/12+o(1)}$ time under Assumption~\ref{ass:arithmetic}, and $\Ot(N^{7/4})$ combinatorial time.
\end{theorem}
\paragraph{Global reduction.} Add/set maps compose into add/set maps, so Lemma~\ref{lem:gap} gives an exact trellised history of length $O(k_C)$ per leaf. Add a local sum query immediately before each partial assignment, on its assigned strip. Solve these and the original partial queries offline. Replay the original global sequence with lazy full-node sums: a partial addition of $a$ on strip $R$ increments the leaf sum by $a|R|$; a partial assignment of $w$ increments it by $w|R|-S_R$, where $S_R$ is its precomputed previous sum. The induction on operations is exact, since all full maps act directly on sums and the partial corrections are exact. Sum full-node and precomputed partial answers for each query. Thus only the local history algorithm remains, with no global temporal batching.

\begin{lemma}[Chronological join]\label{lem:chrono}
A 2D trellised add/set history of $k$ operations, including arbitrary local rectangle sum queries, reduces in $\Ot(k)$ time and size to masked weighted three-matrix products, two-matrix products and one-dimensional historical range sums.
\end{lemma}
\begin{proof}
For a query at time $t$, let $(s_x,w_x)$ and $(s_y,w_y)$ be the latest assignment timestamps and values at $x,y$ on their respective axes (initially zero). Let $A_i(z,u)$ be the cumulative additions covering coordinate $z$ on axis $i$ through time $u$, without regard to assignments. The last writer gives
\begin{equation}\label{eq:last}
V(x,y,t)= A_x(x,t)+A_y(y,t)+\begin{cases}
w_x-A_x(x,s_x)-A_y(y,s_x),&s_x\ge s_y,\\
w_y-A_y(y,s_y)-A_x(x,s_y),&s_y>s_x.
\end{cases}
\end{equation}
On the x-winning region expand $A_y(y,s_x)$ as
\[
A_y(y,s_y)+\sum_{e\text{ y-add}:s_y<u_e\le s_x}a_e\ind_{y\in G_e},
\]
where $(u_e,a_e,G_e)$ are an addition's time, value and interval. The inequality already forces x to win. Use the symmetric expansion on the y-winning region. The two base integrals of $A_i(z,t)$ over a query rectangle are ordinary one-dimensional historical range sums. Any such requested integral is obtained by a time sweep with a range-add/range-sum tree.

\emph{Signed bags.} Each axis's latest assignments partition it into intervals with constant $(s,w)$. An assignment splits at most two boundaries, deletes covered intervals and creates one new interval. Over the entire history only $O(k)$ intervals are created or removed; do not split for additions. Record each creation/deletion with sign $\delta=\pm1$, decomposing its interval into canonical spatial fragments $F$. Copy records to spatial ancestors and into canonical time-prefix subsets of each ancestor list. This is $\Ot(k)$ records in total. For a canonical query interval $I$ at time $t$, if an active fragment contains $I$, use a private record on $I$ with that assignment label. Otherwise use the time prefix of the list at $I$: by laminarity, all intersecting active fragments are descendants. The active-fragment containment test uses a dictionary and ancestor search in the chronological sweep. Decomposing each query rectangle spatially yields polylogarithmically many pairs of bags; their signed sums represent exactly the two current assignment partitions, including the uniform-ancestor case.

For a fragment $F$ on axis $i$, compute
\[
m_F=\delta |F|,\qquad b_F=\delta\int_F A_i(z,s)\,dz,\qquad c_F=\delta w|F|-b_F.
\]
The historical integrals are evaluated by the same sweep. For each bag pair, the two-way correction in~\eqref{eq:last} is the sum of $c_xm_y-m_xb_y$ over $s_x\ge s_y$, plus $m_xc_y-b_xm_y$ over $s_y>s_x$. The comparison labels encode these as a constant number of masked dot products, of total near-linear size.

\emph{Three-way correction.} Decompose every y-add interval into canonical fragments $G$. For the x-winning region, subtract the sum of terms
\begin{equation}\label{eq:triple}
m_x\delta_y a_e |F_y\cap G|\ind_{s_y<u_e\le s_x}.
\end{equation}
Use independent temporal labels $\ell$ for $u_e\le s_x$ and $r$ for $s_y<u_e$. The left outer vertices are x-bags, the right outer vertices y-bags, and the middle vertices are $\ell$ and $(r,H,\eta)$, with spatial node $H$ and case tag $\eta\in\{A,B\}$. Insert the following edges, summing parallel weights:
\begin{center}\small
\begin{tabular}{@{}lll@{}}\toprule
Edge & Condition & Weight\\\midrule
$x$-bag $\to\ell$ & $s_x$ in larger/equal role & $m_x$\\
$\ell\to(r,G,A)$ & $u_e$ in both roles & $a_e$\\
$\ell\to(r,H,B)$ & same; $H$ strict ancestor of $G$ & $a_e|G|$\\
$(r,H,A)\to y$-bag & $s_y$ smaller; $H\supseteq F_y$ & $m_y$\\
$(r,F_y,B)\to y$-bag & $s_y$ smaller & $\delta_y$\\\bottomrule
\end{tabular}
\end{center}
The temporal labels match uniquely iff both comparisons hold. Canonical intervals are laminar: case A handles $F_y\subseteq G$; case B handles $G\subsetneq F_y$; disjoint fragments do not match. Thus the product of the three edge weights gives exactly~\eqref{eq:triple}, including the correct intersection length and sign. Each fragment contributes only polylogarithmically many edges. No query-time filter on the middle edges is needed: surviving assignments have $s_x\le t$, so $u_e\le s_x$ already enforces it. Inactive records cancel by multilinearity. Request the entries corresponding to the query bag pairs. Swap axes for the y-winning correction. This proves the claimed size and correctness.
\end{proof}

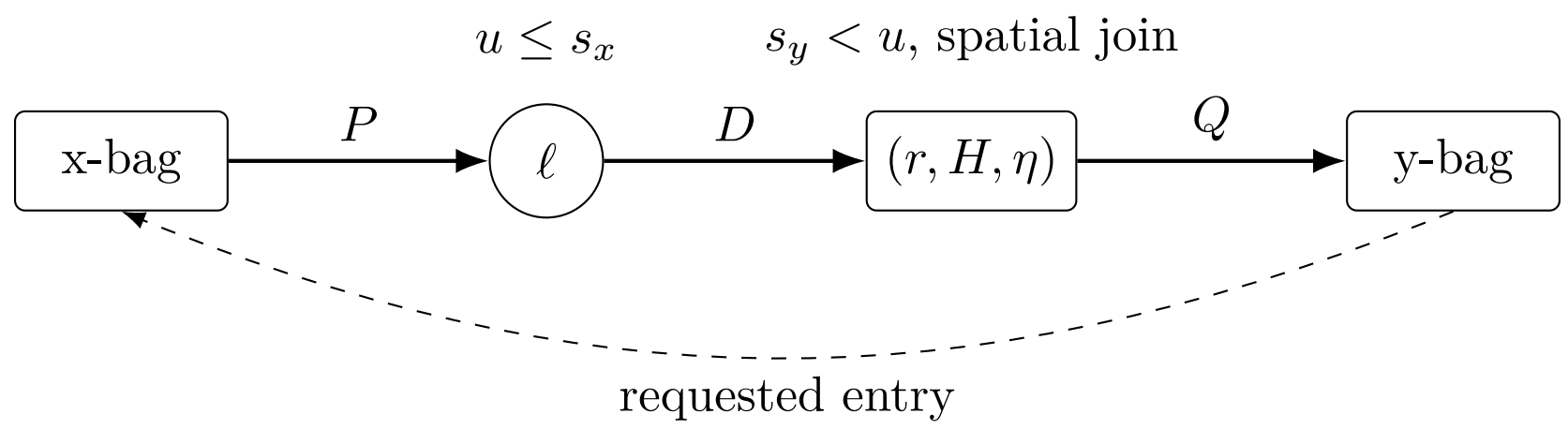
\begin{figure}[ht]
\centering
\begin{tikzpicture}[font=\small,>=Latex]
\node[draw,rounded corners=2pt,minimum width=1.5cm,minimum height=0.7cm] (X) at (0,0) {x-bag};
\node[draw,circle,minimum size=0.8cm] (L) at (3,0) {$\ell$};
\node[draw,rounded corners=2pt,minimum height=0.7cm] (R) at (6,0) {$(r,H,\eta)$};
\node[draw,rounded corners=2pt,minimum width=1.5cm,minimum height=0.7cm] (Y) at (9.4,0) {y-bag};
\draw[->,thick] (X)--node[above]{$P$} (L);
\draw[->,thick] (L)--node[above]{$D$} (R);
\draw[->,thick] (R)--node[above]{$Q$} (Y);
\draw[->,dashed] (Y.south) to[bend left=22] node[below]{requested entry} (X.south);
\node[align=center] at (3,0.85) {$u\le s_x$};
\node[align=center] at (6,0.85) {$s_y<u$, spatial join};
\end{tikzpicture}
\caption{The chronological correction is a weighted masked three-walk sum. The middle matrix is shared across all query times.}
\end{figure}

\begin{proof}[Proof of Theorem~\ref{thm:mixed}]
Apply Lemma~\ref{lem:chrono} and Lemma~\ref{lem:products}, inserting an identity layer if necessary for the two-way terms. The local exponent is $c=(13-\sqrt{19})/6$ or $3/2$ combinatorially. Lemmas~\ref{lem:space}--\ref{lem:gap} and the aggregate replay give total exponent $1+c/2$, as claimed. The previous framework in~\cite{lr}, Theorems 32, 43 and 44 (full version), uses an outer temporal overhead of $N^{1/4}$: the three-walk bound at threshold $\Delta=m^x$ is $\Ot((m+q)\Delta)$ plus the rectangular product of Lemma~\ref{lem:products}, for $m$ graph edges and $q$ queries. At the same $x=(7-\sqrt{19})/6$ this is $\Ot(m^x(m+q))$ with exponent slack, giving the previous global exponent $5/4+c/2$. Our improvement removes that overhead.
\end{proof}

\section{Online monotone updates}\label{sec:online}
\begin{theorem}\label{thm:online}
Fully online Grid Range $(\sum,\chmax)$ admits deterministic combinatorial $\Ot(N^{d/2+3/4})$ total time for fixed $d\ge2$, hence $\Ot(N^{7/4})$ in 2D. The same holds for $\chmin$ and $\set$ separately. For $d>2$ the geometry uses the online partition lifetime guarantee in~\cite{lr}.
\end{theorem}
\paragraph{Online local evaluation.} Use the fragment invariant and intrinsic histograms of Section~\ref{sec:mono}, with implicit axis trees over the bounded integer universe. New nodes without descendants have only intrinsic zero mass. A leaf with $O(B)$ partial operations over its lifetime produces $\Ot(B)$ histogram changes and entries, including initialization and query-created nodes.

For a histogram $Y$, let $G_Y(u)=\sum_w\max(u,w)Y(w)$, evaluated by prefix masses and moments in logarithmic time. Promote an axis histogram permanently when it first reaches $D=\lceil\sqrt B\rceil$ nonzero keys. There are $\Ot(B/D)$ promotions. For every heavy $H_v$ maintain $S_v=\sum_u H_v(u)G_Y(u)$ with $Y$ the opposite root. A mass change $\delta$ at $u$ in $H_v$ changes $S_v$ by $\delta G_Y(u)$; a change at $w$ in $Y$ changes it by $\delta G_{H_v}(w)$. Process changes sequentially. The root pair is always maintained. Initialization and all heavy-cache updates cost $\Ot(B^2/D)$.

In a strip query, the threshold $t$ in~\eqref{eq:phi} is strictly below every positive intrinsic weight of the queried histogram, so
\[
\Phi(H_v,Y,t)=S_v+H_v(0)\sum_{w<t}(t-w)Y(w).
\]
For heavy histograms this is a prefix query; for light ones scan their at most $D$ keys and evaluate $G_Y(\max(t,u))$. Total query work is $\Ot(BD)$, giving $\Ot(B^{3/2})$ per lifetime.

\paragraph{Online geometry and rebuilding.} Fix $B=\lceil\sqrt N\rceil$ for a known horizon. Maintain the x-slabs and global y-quantile intervals of Lemma~\ref{lem:space} by median splits whenever their interior side-occurrence counts exceed $2B$. Children have at most $B+O(1)$ interior occurrences; there are $O(N/B)$ splits of each type. Maintain additional local y-cuts at the endpoints of rectangles with an interior vertical side. An ordinary operation creates $O(1)$ local cuts, a global y-split creates $O(N/B)$ new leaves, and an x-split creates $O(N/B+B)$ leaves. Thus the total number of leaf lifetimes is $O(N+(N/B)^2+(N/B)(N/B+B))=O(N)$. Each leaf has $O(B)$ partial operations, including previous ones. Its past partial operations are found by scanning its interior x- and y-side occurrences, deduplicating and testing intersections.

To reconstruct a newly created leaf, find the maximum $b$ of earlier updates that fully cover it using an insertion-only four-dimensional dominance maximum index on points $(l_x,-r_x,l_y,-r_y)$, with query at the corresponding leaf corner tuple. Such an index is a polylogarithmic-cost dyadic range tree. Rebuild the leaf using the baseline $b$ and its $O(B)$ old partial strips, discarding dominated fragments. This works because maxima commute. Insert new geometry before applying the current operation, so reconstruction uses only earlier updates.

Maintain balanced y-trees within slabs. Before a rotation or leaf replacement, push affected spatial tags to children (an existing child tag is larger and takes precedence). Descendant tags already exceed the pushed weight. This costs polylogarithmic work per structural modification, preserves strict domination, and permits the full-sum recurrence. At a new leaf retain $b$ as its baseline unless a retained ancestor already has that weight; no retained ancestor can exceed $b$. New spatial tags from pushes are charged to structural modifications, all other deletions to prior insertions. Hence the total costs are $\Ot(N^{3/2})$ for geometric pieces and spatial work, and $\Ot(NB^{3/2})=\Ot(N^{7/4})$ for local histories. Unknown horizons are handled by doubling and replaying previous updates at each rebuild.

\paragraph{Fixed dimension.} Use the online level partition of~\cite{lr} (full version, Theorems 21--22): $O(N^{d/2})$ trellised leaf lifetimes, each with $O(B)$ partial boxes. As in Lemma~\ref{lem:space}, binary trees on last-axis children give $\Ot(N^{(d+1)/2})$ pieces; reconstruct using a $2d$-dimensional dominance index. Only the geometric guarantee is imported, not the distributivity requirement of their aggregate algorithm. The uncovered volume is the product of zero masses. For $d$ histograms use a weight tree storing, for each nonempty subset $T\subseteq[d]$, the restricted mass-product maximum sum $M_T$. At a singleton $w$, $M_T=w\prod_{i\in T}m_i(w)$; merging lower and higher ranges gives
\[
M_T=M_{T,L}+\sum_{\emptyset\ne U\subseteq T}M_{U,R}\prod_{i\in T\setminus U}m_{i,L}.
\]
The threshold correction is $t\prod_{i\in T}m_i(\le t)-M_{T,\le t}$. Maintain the root combination, all combinations omitting one axis, and one combined tree per heavy histogram with the other roots. A light histogram is scanned against the omitted-root tree; each opposite-root mass change touches $\Ot(B/D)$ heavy trees. Each promotion builds a tree of size $\Ot(B)$, so local work remains $\Ot(BD+B^2/D)$. Multiplying by the lifetime bound proves Theorem~\ref{thm:online}. The negation and timestamp reductions of Theorem~\ref{thm:mono} preserve online operation order.

\section{Local equivalences and conditional optimality}
\begin{theorem}\label{thm:equiv}
Offline trellised 2D sum/$\chmax$ and sum/$\{\add,\set\}$ with strip queries are equivalent, under near-linear reductions, to weighted masked $AB$ and $PDQ$, respectively. Under OMv, online trellised sum/$\chmax$ has no $O(k^{3/2-\varepsilon})$ total-time algorithm for any constant $\varepsilon>0$.
\end{theorem}
\begin{proof}
The forward reductions are Lemmas~\ref{lem:hist} and~\ref{lem:chrono} with the fragment preprocessing. We give both reverse reductions; signed length coefficients can be handled by splitting each outer operand into positive and negative parts and combining four instances.

For $AB$ with nonnegative entries, use the identity
\[
\max(u,w+1)-2\max(u,w)+\max(u,w-1)=\ind_{u=w}\qquad(u,w\in\mathbb Z).
\]
For each row $i$, form a contiguous x-block $I_i$ with subintervals of lengths $A_{ik}$ and assign each subinterval weight $w_k=3k$ by a strip $\chmax$. Make three copies of the grid. For each column $j$, reserve fresh disjoint y-intervals of lengths $B_{kj}$. Query every requested row strip before and after activating these intervals with weights $w_k+1,w_k,w_k-1$ in the three copies. The finite difference of the three sums, subtracted before from after, gives exactly $(AB)_{ij}$. Each nonzero causes one strip per copy and each requested pair causes two queries per copy. Unused y-coordinates and previous columns cancel. Polynomial weights give polynomial lengths; padding to a square grid does not affect the differences.

For $PDQ$, take $P,Q$ nonnegative and $D$ signed. Form x-intervals $X_{i\ell}$ of lengths $P_{i\ell}$, grouped contiguously by $i$, and y-intervals $Y_{rj}$ of lengths $Q_{rj}$, grouped by $r$. In increasing order of $\ell$, set all x-strips $X_{i\ell}$ to zero; then on the y-group $r$ add $D_{\ell r}-D_{\ell+1,r}$, with a final zero row. Differences have total sparsity $O(\operatorname{nnz}D)$. By telescoping, the final value on $X_{i\ell}\times Y_{rj}$ is $D_{\ell r}$. For each $j$, query each requested x-group $i$ before and after setting all strips $Y_{rj}$ to zero. The difference is $(PDQ)_{ij}$; other columns are disjoint. Both reductions have size linear in explicit nonzeros and mask size, up to logarithmic overhead. A masked product may have arbitrarily many unrequested nonzeros, so output-sensitive fully sparse multiplication~\cite{fmm} does not automatically improve these bounds.

For the online lower bound, take an $n\times n$ Boolean matrix of OMv~\cite{omv} as $A$ in the first reduction. Each of the $n$ arriving vectors activates a new y-block of $n$ coordinates, one for each nonzero vector entry, and queries all $n$ row strips. Subtract the saved finite differences from the previous round to recover the integer products, then test positivity. All answers precede the next vector; coordinates and future block geometry are known in advance. The operation count is $O(n^2)$ including setup. A truly sub-$k^{3/2}$ online algorithm gives truly subcubic OMv, contradicting the conjecture. The lower bound applies to the trellised local problem.
\end{proof}

\section{Static queries via separable profiles}\label{sec:static}
For static queries, all updates precede all queries. We use Lemmas~\ref{lem:space} and~\ref{lem:gap}, including the final full-cover gap. In a trellised leaf, write $V(x_1,\ldots,x_d)$ for its final values. Each strip query asks for an aggregate over an interval of a single marginal profile. We compute those profiles without constructing the $d$-dimensional array.

\begin{theorem}\label{thm:staticsum}
Static $d$D Grid Range $(\sum,\{\add,\set\})$ admits deterministic combinatorial $\Ot(N^{(d+1)/2})$ time. In 2D, $U$ updates and $Q$ queries admit $\Ot((U+Q+1)\sqrt{U+1})$ time.
\end{theorem}
\begin{proof}
Let $s_i(x)$ be the final latest assignment time at $x$ on axis $i$ (zero initially). For event $e$ of time $t_e$ and weight $w_e$ on axis $j$, let $J_i(e)$ be its interval if $i=j$ and the whole axis otherwise. Define
\[
B_i(e)=\begin{cases}
J_i(e)\cap\{x:s_i(x)=t_e\},&e\text{ is set and }i=j,\\
J_i(e)\cap\{x:s_i(x)<t_e\},&\text{otherwise}.
\end{cases}
\]
An add survives exactly if no later assignment covers the point; a set survives exactly if it is the last writer. Consequently,
\begin{equation}\label{eq:sumprofile}
V(x)=\sum_e w_e\prod_i\ind_{x_i\in B_i(e)},\qquad
P_r(x_r)=\sum_e w_e\ind_{x_r\in B_r(e)}\prod_{i\ne r}|B_i(e)|.
\end{equation}
Here $P_r$ is the sum over all coordinates other than $r$. To construct it with $k$ events, compute the latest assignments on the $O(k)$ compressed atoms. All lengths $|B_i(e)|$ are orthogonal counts on (atom position, final timestamp), weighted by physical length; a sweep and Fenwick tree give all counts in $\Ot(k)$ time for fixed $d$.

For fixed $r$, put $c_e=w_e\prod_{i\ne r}|B_i(e)|$. Sweep time backwards with a range-add/point-query tree. At time $s$ read each atom with $s_r=s$ before inserting strict-type events of time $s$ as range additions of $c_e$ on $J_r(e)$. The read includes exactly events with $t_e>s$. Add the equality-type coefficient of the atom's last own-axis assignment, if any. This gives~\eqref{eq:sumprofile}; previously read atoms need not be frozen. Prefix integrals of the profiles answer all strip sums. All preprocessing is $\Ot(k)$.

Add/set maps are closed under composition, so gap compression and final aggregation over the spatial tree yield $\Ot(S)=\Ot(N^{(d+1)/2})$. For the refined 2D bound, partition using updates only, with $B=\lceil\sqrt{U+1}\rceil$, and build the profiles. Each new query has $\Ot(B)$ pieces, and at most four non-strip leaf intersections (each contains a query corner). For any exceptional intersection $I\times J$, evaluate $\sum_e w_e|B_x(e)\cap I||B_y(e)\cap J|$ using the same weighted count sweep in $\Ot(B)$ time. Strip queries use prefix integrals, including truncation of boundary atoms. Thus the cost is $\Ot((U+1)^{3/2}+QB)$.
\end{proof}

\begin{theorem}\label{thm:staticmax}
Static $d$D Grid Range $(\max,\{\add,\set,\chmax\})$ and $(\max,\{\set,\chmin,\chmax\})$ admit deterministic combinatorial $\Ot(N^{(d+1)/2})$ time for fixed $d\ge2$, including every subset of either family.
\end{theorem}
\begin{proof}
\emph{Add/set/chmax.} Replace each assignment $w$ globally by an addition $-4W$ followed by $\chmax_w$, where $W=1+$ the sum of absolute original weights. At each original operation boundary values lie in $[-W,W]$, so the replacement is exact. The resulting full-cover maps $f_{a,b}(v)=\max(v+a,b)$ compose according to
\[
f_{c,h}\circ f_{a,b}=f_{a+c,\max(b+c,h)}.
\]
They are represented by at most two updates (allow $b=-\infty$). Lemma~\ref{lem:gap} gives a trellised history of length $O(k_C)$.

Regard each $\chmax$ event $e$ and the initial zero as a source of time $s_e$, weight $w_e$ and separable support $\prod_i J_i(e)$, with only one restricted axis. Define the suffix-addition function $D_i(x,s)=\sum_{f\text{ add on }i:t_f>s} a_f\ind_{x\in J_i(f)}$. Inducting over the update sequence (add translates all applicable sources, chmax introduces a source) gives
\[
V(x)=\max_{e:x_i\in J_i(e)\ \forall i}\left(w_e+\sum_i D_i(x_i,s_e)\right).
\]
For every source compute $M_i(e)=\max_{x\in J_i(e)}D_i(x,s_e)$ by a backward range-add/range-max sweep on each axis. Maximizing over the other coordinates now yields
\begin{equation}\label{eq:maxprofile}
P_r(x)=\max_{e:x\in J_r(e)}\left(w_e+\sum_{i\ne r}M_i(e)+D_r(x,s_e)\right).
\end{equation}
To construct this profile, sweep forwards on axis $r$, inserting each source by $\chmax$ of $w_e+\sum_{i\ne r}M_i(e)$ on $J_r(e)$ and applying actual additions on axis $r$. Initialize at $-\infty$ before inserting the initial source. The maps $f_{a,b}$ act lazily on range maxima and are closed under composition. Thus~\eqref{eq:maxprofile} is computed in $\Ot(k_C)$ per leaf. Read strip and full-leaf maxima, aggregate full nodes bottom-up, and use the canonical query pieces. Total time is $\Ot(S)$.

\emph{Set/chmin/chmax.} Full-cover clamp maps $v\mapsto\min(\max(v,l),h)$ compose into clamps, allowing infinite endpoints and constants. Compress their gaps as above. For a fixed threshold $z$, translate events into Boolean writes for the predicate $V\ge z$: $\chmax_w$ writes true if $w\ge z$, $\chmin_w$ writes false if $w<z$, and set writes its comparison with $z$. Other events do nothing. Give true and false writes signed labels $+t,-t$ at distinct times $t\ge2$, and initialize with $+1$ or $-1$ according to $0\ge z$. The latest label is the one of largest absolute value. Its sign determines the predicate. Increasing any signed label numerically cannot turn true into false: a positive label becomes later, a negative label becomes earlier, or negative becomes positive. Thus the maximum predicate over a box is obtained by independently maximizing the numerical latest label over each axis interval.

For each axis, use a segment tree with multisets of active write labels at canonical nodes. Let $L(a,b)$ return the label of larger absolute value. A node with latest local tag $t$ and children's numerical maxima $A_1,A_2$ has intrinsic maximum $L(t,\max(A_1,A_2))$, since $L(t,\cdot)$ is nondecreasing. Empty tags are identities, and elementary leaves carry the initial label. Interval label insertion/deletion and query with inherited tags take polylogarithmic time. Sweep $z$ upwards: each chmax label is removed once, chmin inserted once, and set changes from positive to negative once. The initial label changes at zero. Parallel binary search on integer answer values, sorting trial thresholds each round, gives all local static query maxima in $\Ot(k_C)$ time, even for arbitrary local boxes. This adapts the signed-label method of~\cite{lr}. Aggregate as above.
\end{proof}

\begin{corollary}[Static maximum exponent gap]\label{cor:gap}
For fixed $d\ge2$, consider static maximum queries with update family $\mathcal U\subseteq\{\add,\set,\chmin,\chmax\}$. The problem admits $\Ot(N^{(d+1)/2})$ time unless $\mathcal U$ contains both add and chmin. In the latter case the $\Ot(N^d)$ baseline is conditionally optimal up to subpolynomial factors under $d$-OV.
\end{corollary}
\begin{proof}
Every family not containing both add and chmin lies in one of the families of Theorem~\ref{thm:staticmax}. The hardness is Lau--Ritossa~\cite{lr}, full version Lemma 17, even for a single final query on a trellis. For the baseline, compress all endpoints and maintain a lazy 1D range tree for every $(d-1)$-tuple of atoms on the other axes. Each of the $O(N)$ operations costs $\Ot(N^{d-1})$. All four update maps are monotone and compose into a translated clamp (allowing constant maps), so lazy maximum maintenance is valid.
\end{proof}

\begin{proposition}[Weighted clique transfer]\label{prop:clique}
Maximum edge-weight $(d+1)$-partite clique on $n$ vertices per part reduces to static $d$D Grid Range $(\max,\{\add,\set\})$ using $O_d(n^2)$ updates and one final query.
\end{proposition}
\begin{proof}
Index axis 1 by pairs $(a,b)$ in lexicographic order (with $a$ first), and axes $j\ge2$ by $c_j$. For $b=1,\ldots,n$, assign $W_{ab}(a,b)$ on each axis-1 atom $(a,b)$; then add $W_{bc_j}(b,c)-W_{bc_j}(b+1,c)$ to every corresponding axis-$j$ strip, using a fictitious zero row at $b=n+1$. Telescoping after the last assignment gives $W_{ab}(a,b)+\sum_j W_{bc_j}(b,c_j)$. Add each remaining edge weight $W_{ac_j}(a,c)$ on the box fixing the contiguous $a$-group and coordinate $c$, and each $W_{c_i c_j}$ on the box fixing those two coordinates. The final value is the clique weight. Query the maximum on the used domain, excluding padding. There are $O_d(n^2)$ operations with signed polynomial weights. Nonpartite inputs can use copies of the vertex set and sufficiently negative forbidden-edge penalties.
\end{proof}
This transfers a static exponent $c$ to a weighted-clique exponent $2c$. The $N^{(d+1)/2}$ upper bound is therefore tight under the corresponding $n^{d+1-o(1)}$ weighted-clique hypothesis.

\section{Discussion}
The two dynamic sum algorithms share the same spatial accounting but rely on distinct local identities: strict domination for monotone updates and last-writer cancellation for assignments. The remaining major question is dynamic maximum with add and set. The sum replay of Section~\ref{sec:mixed} does not apply to maxima: a partial assignment's old and new maxima do not determine the maximum of its complement. Static profiles do not remove this difficulty at intermediate times. A subquadratic local algorithm with efficient full-region maintenance would yield a subquadratic global bound. Likewise, the OMv result constrains the online local primitive, not the unrestricted geometric exponent.

\paragraph{Use of generative AI.} Generative AI tools assisted with literature searches, drafting and editing. The authors take responsibility for all mathematical claims and conclusions.

\begin{thebibliography}{9}
\bibitem{lr} Joshua Lau and Angus Ritossa. Algorithms and hardness for multidimensional range updates and queries. \emph{ITCS 2021}, LIPIcs 185, 35:1--35:20. \url{https://doi.org/10.4230/LIPIcs.ITCS.2021.35}. Full version: \url{https://arxiv.org/abs/2101.02003}.
\bibitem{oy} Mark H. Overmars and Chee-Keng Yap. New upper bounds in Klee's measure problem. \emph{SIAM J. Comput.}, 20(6):1034--1045, 1991.
\bibitem{uz} Przemys\l aw Uzna\'nski. Rectangular matrix multiplication from shared-leg entropy. \url{https://arxiv.org/abs/2610.10494v1}, 2026.
\bibitem{bh} Paul Burkhardt and David G. Harris. Simple and efficient four-cycle counting on sparse graphs. \url{https://arxiv.org/abs/2303.06090}, 2023.
\bibitem{omv} Monika Henzinger, Sebastian Krinninger, Danupon Nanongkai, and Thatchaphol Saranurak. Unifying and strengthening hardness for dynamic problems via the online matrix-vector multiplication conjecture. \emph{STOC 2015}. \url{https://arxiv.org/abs/1511.06773}.
\bibitem{bs} Walter Baur and Volker Strassen. The complexity of partial derivatives. \emph{Theor. Comput. Sci.}, 22(3):317--330, 1983.
\bibitem{fmm} Amir Abboud, Karl Bringmann, Nick Fischer, and Marvin K\"unnemann. Fully sparse matrix multiplication. \emph{SODA 2024}. \url{https://arxiv.org/abs/2309.06317}.
\end{thebibliography}
\end{document}